\documentclass[aps,prb,eqsecnum,twocolumn,citeautoscript,floatfix]{revtex4}
\usepackage{pstricks,pstcol}
\usepackage{amsmath,amssymb,braket,dcolumn}
\usepackage{graphicx,mathrsfs,textcomp,wasysym}
\usepackage{bbold}
\usepackage{txfonts}
\newcommand{\figref}[1]{Fig.\,\ref{#1}}
\newcommand{\cx}[1]{{c_{#1}^{\phantom{\dagger}}}}
\newcommand{\cxd}[1]{{c_{#1}^{\dagger}}}

\newcommand{\nbar}{\ensuremath{\bar{n}}}

\newcommand{\im}{\mathrm{i}}
\newcommand{\bR}{\mathbf{R}}
\newcommand{\bQ}{\mathbf{Q}}
\newcommand{\bi}{\mathbf{i}}
\newcommand{\bj}{\mathbf{j}}
\newcommand{\bk}{\mathbf{k}}
\newcommand{\bl}{\mathbf{l}}
\newcommand{\bm}{\mathbf{m}}
\newcommand{\bn}{\mathbf{n}}

\newcommand{\bphi}{\boldsymbol{\phi}}

\DeclareMathOperator{\Tr}{Tr}
\psset{arrowsize=5pt}
\begin{document}
\title{Many-Body Density Matrices On a Two-Dimensional Square
Lattice:\\ Noninteracting and Strongly Interacting Spinless Fermions}
\author{Siew-Ann \surname{Cheong}}
\affiliation{Cornell Theory Center, Cornell University,
Ithaca, New York 14853-3801, USA}
\author{Christopher L. \surname{Henley}}
\affiliation{Laboratory of Atomic and Solid State Physics, Cornell University,
Ithaca, New York 14853-2501, USA}
\date{\today}

\begin{abstract}
\bigskip
The reduced density matrix of an interacting system can be used as the 
basis for a truncation scheme, or in an unbiased method to discover the
strongest kind of correlation in the ground state.  In this paper, we 
investigate the structure of the many-body fermion density matrix of a 
small cluster in a square lattice.  The cluster density matrix is 
evaluated numerically over a set of finite systems, subject to non-%
square periodic boundary conditions given by the lattice vectors
$\bR_1 \equiv (R_{1x}, R_{1y})$ and $\bR_2 \equiv (R_{2x}, R_{2y})$.
We then approximate the infinite-system cluster density-matrix 
spectrum, by averaging the finite-system cluster density matrix (i) 
over degeneracies in the ground state, and orientations of the system 
relative to the cluster, to ensure it has the proper point-group 
symmetry; and (ii) over various twist boundary conditions to reduce 
finite size effects.  We then compare the eigenvalue structure of the 
averaged cluster density matrix for noninteracting and strongly-%
interacting spinless fermions, as a function of the filling fraction
$\nbar$, and discuss whether it can be approximated as being built up
from a truncated set of single-particle operators.
\end{abstract}
\maketitle

\section{Introduction}
\label{sect:intro}

The density matrix is a very useful tool in the numerical study of
interacting systems.  Besides being used in the Density-Matrix 
Renormalization Group (DMRG)\cite{white92} and its higher-dimensional
generalizations,\cite{verstraete04} the density matrix is also used as
a diagnostic tool in the Contractor Renormalization (CORE) method for
numerical renormalization group in two dimensions,\cite{core}
and forms the basis of a method to identify the order parameter related
to a quasi-degeneracy of ground states.\cite{furukawa05}

In previous work,\cite{cheong04a} we extended the results of Chung and 
Peschel\cite{chung01} to write the density matrix (DM) of a cluster of
$N_C$ sites cut out from a system of noninteracting spinless fermions in
$d$ dimensions as the exponential of a quadratic operator, called the 
pseudo-Hamiltonian, as it resembles the Hamiltonian of a noninteracting 
system.  That result was then applied in numerical studies of noninteracting 
spinless fermions in one dimension, to better understand how the 
distribution of cluster DM eigenvalues scale with $N_C$, and to explore
the possibility of designing truncation schemes based on the pseudo-%
Hamiltonian.\cite{cheong04b}  We believe truncation schemes such as that
described in Ref.~\onlinecite{cheong04b} will be helpful to the choice
of basis states in renormalization groups such as CORE.

Thus, some questions motivating the present paper were: (i) does the
density matrix of an interacting Fermi-liquid system resembles that of
a noninteracting one? (ii) can we apply our exact result in 
Ref.~\onlinecite{cheong04a} to two dimensions as well as for one dimension?
(iii) is it numerically practical to compute this sort of density matrix
in a fermion system.  To answer these questions, we investigated a spinless 
analog of the extended Hubbard model, given by the Hamiltonian
\begin{equation}\label{eqn:strongintham}
H = -t\sum_{\braket{i, j}}\cxd{i}\cx{j} + V\sum_{\braket{i, j}} n_i n_j,
\end{equation}
in the limit of $V \to \infty$, so that fermions are not allowed to be
nearest neighbors of each other.  This model is chosen for two
reasons: (i) for a given number of particles, the $V \to \infty$ 
Hilbert space is significantly smaller than the $V < \infty$ Hilbert 
space, and we can work numerically with larger system sizes; and (ii)
the model, in spite of its simplicity, has a rich zero-temperature 
phase diagram,\cite{zhang01,zhang03,zhang04} where we find practically 
free fermions in the limit $\nbar \ll 1$, and an inert solid at half-%
filling $\nbar = \frac{1}{2}$.  As the filling fraction approaches 
quarter-filling from below, $\nbar \to \frac{1}{4}^{-}$, the system
becomes congested, highly correlated, but is nonetheless a Fermi liquid, 
perhaps with additional orders that are not clear in small systems.  
Slightly above quarter-filling, the dense fluid and inert solid coexists,
while slightly below half-filling, the system is expected to support
stable arrays of stripes.

To probe this rich variety of structures in the ground state at different 
filling fraction $\nbar$, we describe in Section~\ref{sect:formulation} 
how the reduced DM of a small cluster, with the appropriate symmetry 
properties, can be calculated from a finite non-square system subject to
twist boundary conditions.  Then in Section~\ref{sect:ni}, we investigate
in great details the cluster DM spectra of the noninteracting system,
particularly on how to handle finite size effects in the numerics, for 
comparison with the cluster DM spectra of a strongly-interacting 
system, presented in Section~\ref{sect:strong}.  Finally, in 
Section~\ref{sect:summary}, we summarized our findings, and discuss the 
prospects of designing an Operator-Based DM Truncation Scheme for 
interacting systems, at some, if not at all, filling fractions.

\section{Formulation}
\label{sect:formulation}

In this section, we give the theoretical formulations and describe
the numerical tools needed to investigate the cluster DM 
spectra of noninteracting and strongly-interacting systems of spinless
fermions in two dimensions.  In Section~\ref{sect:cdmf}, we give the
matrix elements of the DM of a small cluster embedded within a larger,
but still finite, system.  These matrix elements are obtained by 
tracing out degrees of freedom external to the cluster, starting from 
the ground-state wave function of the system, obtained through exact 
diagonalization.  In Section~\ref{sect:sysdef}, we describe how our 
finite systems can be defined with nonsquare periodic boundary 
conditions, and how we make use of the translational invariance of 
both noninteracting and strongly-interacting models to reduce the 
computational efforts in exact diagonalization.  In
Section~\ref{sect:averaging}, we describe several averaging apparatus
required to obtain a handle on the infinite-system spectra of the 
cluster DM, and then in Section~\ref{sect:states}, we describe a 
classification scheme for the one-particle and multi-particle eigenstates 
of the cluster DM that makes the symmetry of the underlying square lattice 
explicit.

\subsection{Cluster Density Matrix}
\label{sect:cdmf}

The DM $\rho_C$ of a cluster cut out from a larger system is a density
operator which gives the expectation
\begin{equation}\label{eqn:definerhoc}
\braket{\Psi|A|\Psi} = \braket{A} = \Tr_C \rho_C A
\end{equation}
for any observable $A$ local to the cluster, when the larger system is
in its ground state $\ket{\Psi}$.  The cluster DM $\rho_C$ can be 
calculated from the ground-state DM
\begin{equation}\label{eqn:groundstatedm}
\rho = \ket{\Psi}\bra{\Psi}
\end{equation}
of the system, by tracing out degrees of freedom outside of the cluster.
We write this as
\begin{equation}\label{eqn:environmenttrace}
\rho_C = \Tr_E \rho,
\end{equation}
where the subscript $E$ denotes a trace over environmental degrees of
freedom.

Since a cluster is a collection of sites identified in real space, it is
natural to choose as a many-body basis the real-space configurations.
For a finite two-dimensional system with $N$ sites, we label the sites 
$j = 1$ through $j = N$, so that for any pair of sites $(x_{j_1}, y_{j_1})$ 
and $(x_{j_2}, y_{j_2})$, we have $x_{j_1} \leq x_{j_2}$ and $y_{j_1} < 
y_{j_2}$ if $j_1 < j_2$.   We then distinguish between sites within the
cluster, of which there are $N_C$ of them, $(x_{j_1^C}, y_{j_1^C})$,
$(x_{j_2^C}, y_{j_2^C})$, \dots, $(x_{j_{N_C}^C}, y_{j_{N_C}^C})$, and
sites outside of the cluster, of which there are $N_E = N - N_C$ of them,
$(x_{j_1^E}, y_{j_1^E})$, $(x_{j_2^E}, y_{j_2^E})$, \dots, $(x_{j_{N_E}^E},
y_{j_{N_E}^E})$.  We think of the $N_E$ sites outside the cluster as
constituting the \emph{environment} to the cluster.

We work with the configuration basis states $\ket{\bj} = \ket{j_1 j_2 \cdots 
j_P}$, where $j_1 < \cdots < j_P$ are the $P$ occupied sites in the system.
These can be thought of as a direct product of the configuration basis
states of the cluster $\ket{\bl} = \ket{l_1 l_2 \cdots l_{P_C}}$, where
$l_1 < \cdots < l_{P_C}$ are the $P_C$ occupied sites within the cluster, 
and the configuration basis states of the environment $\ket{\bm} = \ket{m_1
m_2 \cdots m_{P_E}}$, where $m_1 < \cdots < m_{P_E}$ are the $P_E = P - P_C$ 
occupied sites in the environment.  Here, we have the occupied sites of the
system $\{j_1, \dots, j_P\} = \{l_1, \dots, l_{P_C}\}\cup\{m_1, \dots, 
m_{P_E}\}$ being the union of the occupied sites in the cluster and in the
environment, with the site indices $l$ and $m$ resorted in ascending order
to give the site indices $j$.  

In terms of the configuration basis of the system, the ground-state wave 
function of the system can be written as
\begin{equation}\label{eqn:systemexpansion}
\ket{\Psi} = \sum_{\bj}\Psi_{\bj}\ket{\bj} =
\sum_{\bj}\Psi_{\bj}\cxd{j_1}\cdots\cxd{j_P}\ket{0},
\end{equation}
where $\Psi_{\bj}$ is the amplitude associated with configuration 
$\ket{\bj}$, and $\cx{j}$, $\cxd{j}$ are fermion annihilation and 
creation operators acting on the site $(x_j, y_j)$.  We can also
write the ground-state wave function as
\begin{equation}\label{eqn:directproductexpansion}
\begin{split}
\ket{\Psi} &= \sum_{\bl}\sum_{\bm} (-1)^{f(\bj;\bl,\bm)}
\Psi_{\bl,\bm}\ket{\bl}\ket{\bm} \\
&= \sum_{\bl}\sum_{\bm} (-1)^{f(\bj;\bl,\bm)} \Psi_{\bl,\bm} \times {} \\
&\qquad\qquad\qquad \cxd{m_1} \cdots \cxd{m_{P_E}}
\cxd{l_1} \cdots \cxd{l_{P_C}}\ket{0},
\end{split}
\end{equation}
in terms of the direct product of configuration bases of the cluster and
the environment, where $\cx{l}$ and $\cxd{l}$ are fermion annihilation
and creation operators acting on site $(x_l, y_l)$ within the cluster, and
$\cx{m}$ and $\cxd{m}$ are fermion annihilation and creation operators
acting on site $(x_m, y_m)$ within the environment.  In 
\eqref{eqn:directproductexpansion}, the amplitude $\Psi_{\bl,\bm} = 
\Psi_{\bj}$ is taken directly from the expansion in 
\eqref{eqn:systemexpansion}, while the factor $(-1)^{f(\bj;\bl,\bm)}$ 
accounts for the fermion sign incurred when we
reorder the operator product $\cxd{m_1} \cdots \cxd{m_{P_E}}
\cxd{l_1} \cdots \cxd{l_{P_C}}$ to get the operator product
$\cxd{j_1}\cdots\cxd{j_P}$.

Similarly, the ground-state DM in \eqref{eqn:groundstatedm} can be written as
\begin{equation}
\rho = \sum_{\bj}\sum_{\bj'}\Psi_{\bj}\Psi_{\bj'}^*
\cxd{j_1}\cdots\cxd{j_P}\ket{0}\bra{0}\cx{j'_P}\cdots\cx{j'_1},
\end{equation}
using the system-wide configuration basis, or as
\begin{multline}\label{eqn:directproductgroundstatedm}
\rho = \sum_{\bl, \bm}\sum_{\bl', \bm'}(-1)^{f(\bj;\bl,\bm) + 
f(\bj';\bl',\bm')} \Psi_{\bl,\bm}\Psi_{\bl',\bm'}^* \times {} \\
\cxd{m_1}\cdots\cxd{m_{P_E}}\cxd{l_1}\cdots\cxd{l_{P_C}}\ket{0} \times {} \\
\bra{0} \cx{l'_{P'_C}}\cdots\cx{l'_1}\cx{m'_{P'_E}}\cdots\cx{m'_1},
\end{multline}
using the direct-product basis between cluster configurations and
environment configurations.

Performing the trace over the environment as prescribed in 
\eqref{eqn:environmenttrace}, we find the fermion cluster DM $\rho_C$
to be
\begin{multline}\label{eqn:detailrhoc}
\rho_C = \sum_{\bl,\bl'}\sum_{\bm,\bm'}
(-1)^{f(\bj;\bl,\bm) + f(\bj';\bl',\bm')} \times {} \\
\Psi_{\bl,\bm}\Psi_{\bl',\bm'}^* 
\delta_{\bm,\bm'} \times {} \\
\cxd{l_1}\cdots\cxd{l_{P_C}}\ket{0}\bra{0}
\cx{l'_{P'_C}}\cdots\cx{l'_1}.
\end{multline}
Its matrix elements are
\begin{multline}\label{eqn:detailrhocmatrixelements}
\braket{\bl|\rho_C|\bl'} = \sum_{\bm}\sum_{\bm'}
(-1)^{f(\bj;\bl,\bm) + f(\bj';\bl',\bm')} \times {} \\
\Psi_{\bl,\bm}\Psi_{\bl',\bm'}^* \delta_{\bm,\bm'}.
\end{multline}

These matrix elements can be computed naively by running over all
possible pairs of cluster states $\ket{\bl}$ and $\ket{\bl'}$, and
performing the sums over $\bm$ and $\bm'$ as they appear in
\eqref{eqn:detailrhocmatrixelements}, looking up the fermion-sign
factors $(-1)^{f(\bj;\bl,\bm)}$ and amplitudes $\Psi_{\bl,\bm}$ as and
when they are needed.  We call this the \emph{naive algorithm}.
Alternatively, we can also reorganize the fermion-sign-factor-adjusted 
amplitudes $(-1)^{f(\bj;\bl,\bm)}\Psi_{\bl,\bm}$ into a matrix 
$\tilde{\boldsymbol{\Psi}}$, whose rows are associated with the 
cluster configurations $\ket{\bl}$, and whose columns are associated 
with the environment configurations $\ket{\bm}$, the matrix $\rho_C$ 
can be computed directly by matrix multiplication as
\begin{equation}\label{eqn:presortinnerproduct}
\rho_C = \tilde{\boldsymbol{\Psi}}\tilde{\boldsymbol{\Psi}}^{\dagger},
\end{equation}
representing a collection of inner products.  We call this the
\emph{pre-sorted inner-product algorithm}.  We compare and analyze the
computational complexity of these two algorithms in Appendix 
\ref{appendix:computational}.  In the numerical studies presented in
Section \ref{sect:strong}, we use the pre-sorted inner-product
algorithm exclusively.

From \eqref{eqn:presortinnerproduct}, we see that the cluster DM
$\rho_C$ is manifestly hermitian, and thus all its eigenvalues $\{w\}$
are real (and in fact nonnegative).  When obtained from the wave
function of a state with definite particle number, $\rho_C$ has no
matrix elements between cluster states containing different number of
particles, and thus the eigenstates $\ket{w}$ of $\rho_C$ can be
organized into \emph{sectors}, corresponding to $P_C = 0, 1, \dots,
P_{C,\max}$ particles within the cluster.  For the rest of this paper,
we would refer to the eigenvalues of $\rho_C$ generically as its
\emph{weights}, since these have a natural probabilistic
interpretation.  A $P_C$-particle weight of $\rho_C$ is therefore an
eigenvalue corresponding to an eigenstate containing $P_C$ particles
within the cluster.

\subsection{System Definition and Translational Invariance}
\label{sect:sysdef}

\begin{figure}[htbp]
\centering
\includegraphics{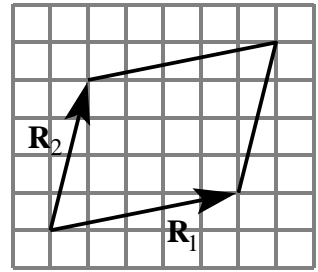}
\caption{Definition of system to be exactly diagonalized in terms of the
lattice vectors $\bR_1$ and $\bR_2$.  We shall denote such a system as
$\bR_1\times\bR_2$.  In this example shown, the system is $(5,1)\times(1,4)$.}
\label{fig:system}
\end{figure}

For noninteracting spinless fermions on an infinite square lattice, it is 
possible to compute the cluster DM $\rho_C$ starting from the Fermi sea 
ground state, through the evaluation and diagonalization of $G_C$.  For an 
interacting system, we need to compute $\rho_C$ starting from $\rho$ in
\eqref{eqn:groundstatedm}, the latter we obtain through exact diagonalization 
on a finite system.  We define the a finite system relative to the infinite 
square lattice in terms of the lattice vectors $\bR_1$ and $\bR_2$, as shown 
in \figref{fig:system}, such that $N = \hat{\mathbf{z}} \cdot(\bR_1\times
\bR_2) = R_{1x}R_{2y} - R_{2x}R_{1y} > 0$ is the number of lattice sites 
within the system.

If we impose periodic boundary condition such that $\bR + m\bR_1 + n\bR_2 
\equiv \bR$, then in the exact diagonalization to obtain $\ket{\Psi}$ we can 
take advantage of translational invariance through the use of the Bloch states
\begin{equation}\label{eqn:blochbasis}
\ket{\bj_0, \bk} = \frac{1}{\sqrt N}\sum_{\bR} e^{-\im\bk\cdot\bR} 
T_{\bR}\ket{\bj_0}
\end{equation}
as our computational basis\cite{zhang03}.  In this Bloch state, the
configurations $\{T_{\bR}\ket{\bj_0}\}$ are all related to the generating
configuration $\ket{\bj_0}$ by the lattice translations $T_{\bR}$ 
associated with displacement $\bR$, while $\bk$ are wave vectors allowed 
by the boundary conditions.  Any configuration within the collection of
translationally-related configurations $\{T_{\bR}\ket{\bj_0}\}$ can serve
as the generating configuration, but we pick the one with the least sum of
indices of occupied sites.

Working with finite non-square systems introduces several complications.  
First of all, we sometimes end up with degenerate ground states which 
suffer from symmetry-breaking not found in the true infinite-system ground 
state.  However, because the point symmetry group of our non-square finite 
system is only a subgroup of the square lattice point symmetry group, the 
finite-system ground-state manifold is not invariant under all square lattice 
symmetry operations. 
Thirdly, when working with finite systems, we introduce systematic deviations 
which are collectively known as \emph{finite size effects}.  We identify the 
three primary sources of finite size effects as (i) finite domain effect, which 
has to do with the fact that the small set of discrete wave vectors allowed 
are not adequately representative of the continuous set of wave vectors on 
the infinite square lattice; (ii) shell effect, which has to do with the fact 
that the set of discrete wave vectors allowed are organized by symmetry into 
shells in reciprocal space, each of which can be partially or fully filled in 
the many-body ground state; and (iii) shape effect, which has to do with
the detailed shape of the non-square system we introduced.

\subsection{Averaging}
\label{sect:averaging}

\subsubsection{Degeneracy Averaging}

To eliminate these numerical artefacts, we adopted three averaging devices.
First, we average over the $D_0$-fold degenerate ground-state manifold.
Our first motivation for doing so is as follows: if $\mathscr{G}$ is the 
point symmetry group of the square lattice, and its subgroup $G$ is the point 
symmetry group of the $\bR_1\times\bR_2$ system, then we will find that the 
cluster density matrices 
\begin{equation}\label{eqn:degenerateclusterdm}
\rho_{C, i} = \Tr_E \ket{\Psi_i}\bra{\Psi_i},
\end{equation}
one for each wave function $\Psi_i$ within the $D_0$-fold degenerate 
ground-state manifold, are not invariant under $G$, much less $\mathscr{G}$.  
We remove this artificial symmetry breaking by calculating the degeneracy-%
averaged cluster DM
\begin{equation}
\rho_C = \frac{1}{D_0}\sum_{i=1}^{D_0}\rho_{C,i}
\end{equation}
over the cluster density matrices $\rho_{C, i}$ within the ground-state
manifold.  This degeneracy-averaged cluster DM is invariant
under $G$.

A second motivation for such a mode of averaging over the degenerate 
ground-state manifold of the finite system is that thermodynamically, 
given the pure state density matrices $\rho_{C,i}$, with energy eigenvalue 
$E_i$, we typically construct the canonical ensemble DM as
\begin{equation}
\rho_C(\beta) = Z^{-1}(\beta) \sum_i e^{-\beta E_i}\, \rho_{C,i},
\end{equation}
where $Z(\beta) = \sum_i e^{-\beta E_i}$ is the canonical partition function.
States within a degenerate manifold have the same energy, and therefore
contribute equally to the thermodynamic DM $\rho_C(\beta)$. 
In the limit of $\beta \to \infty$, the usual thermodynamic argument
is that pure states decouple from one another, and we treat their
respective density matrices independently, except for those states which
are degenerate.  Because they appear with the same Boltzmann weight
whatever the inverse temperature is, we should still treat the uniform 
combination instead of the individual density matrices in the limit of 
$\beta \to \infty$,

\subsubsection{Orientation Averaging}

The second averaging device involves an average over the orientation of
the finite non-square system relative to the underlying square lattice.
This averaging restores the $\mathscr{G}$-symmetry to the averaged
ground state.  In principle, this requires us to compute $\rho_C$ for 
a group of four systems: $(R_{1x}, R_{1y})\times(R_{2x}, R_{2y})$, 
$(R_{2y}, R_{2x})\times(R_{1y}, R_{1x})$, $(R_{1x}, -R_{1y})\times(-R_{2x}, 
R_{2y})$ and $(R_{2y}, -R_{2x})\times(-R_{1y}, R_{1x})$.

However, if the cluster whose DM we are calculating is invariant under
the action of $\mathscr{G}$, this averaging can be achieved by computing
\begin{equation}
\bar{\rho}_C = \frac{1}{D(\mathscr{G})}\sum_{g\in \mathscr{G}} 
U_g \rho_C U_g^{\dagger},
\end{equation}
where $g \in G$ is a point group transformation of the square lattice,
$U_g$ is the unitary transformation of the cluster Hilbert space
associated with $g$, and $D(\mathscr{G})$ is the order of $\mathscr{G}$.

\subsubsection{Twist Boundary Conditions Averaging}

After these two averagings, the cluster DM has the full symmetry 
(including translations) of the underlying square lattice, but finite size 
effects remain.  We eliminate these as much as we can\cite{cheong05} with 
the third averaging device, twist boundary conditions averaging.\cite{twist} 
The usual way to implement twist boundary conditions is to work in the
\emph{boundary gauge}, keeping the Hamiltonian unchanged, and demanding that
\begin{equation}\label{eqn:twistboundaryconditions}
\cx{\bR + \bR_1} = e^{-\mathrm{i}\bphi\cdot\bR_1}\,\cx{\bR}, \quad
\cx{\bR + \bR_2} = e^{-\mathrm{i}\bphi\cdot\bR_2}\,\cx{\bR},
\end{equation}
where $\bR_1$ and $\bR_2$ are the lattice vectors defining our finite
system, $\bR = (R_x, R_y)$ is a site within the system, and $\bphi = (\phi_x,
\phi_y)$ is the twist vector parametrizing the twist boundary conditions.

In choice of gauge \eqref{eqn:twistboundaryconditions}, the Hamiltonian
\eqref{eqn:strongintham} is not manifestly invariant under translations.
However, we can continue to block-diagonalize it using the Bloch basis states
defined in \eqref{eqn:blochbasis}, provided the set of allowed wave vectors 
$\bk$ are shifted relative to $\bk_0$ for the usual periodic boundary 
conditions by the twist vector $\bphi$, i.e.
\begin{equation}
\bk = \bk_0 + \bphi.
\end{equation}

The other natural way to implement twist boundary conditions is in the 
\emph{bond gauge}, where we make the substitution
\begin{equation}
\cx{\bR} \to e^{-\mathrm{i}\bR\cdot\bphi}\cx{\bR}
\end{equation}
in the Hamiltonian, but demand that
\begin{equation}
\cx{\bR + \bR_1} = \cx{\bR}, \quad 
\cx{\bR + \bR_2} = \cx{\bR},
\end{equation}
where $\bR_1$ and $\bR_2$ are the lattice vectors defining our finite system.
Now the Hamiltonian \eqref{eqn:strongintham} is manifestly invariant under
translations, and we can bloch-diagonalize it using the Bloch basis states
defined in \eqref{eqn:blochbasis}, with the same set of allowed wave vectors 
$\bk = \bk_0$ as for the usual periodic boundary conditions.

Exact diagonalization can be performed in any gauge, but we chose to do it in
the bond gauge, because in this gauge, the Bloch basis states $\ket{\bj_0,
\bk}$ defined in \eqref{eqn:blochbasis} can be used as is to block diagonalize 
the Hamiltonian at all twist vectors $\bphi$.  This gives us the ground-state 
wave function $\ket{\Psi(\text{bond})}$ in the bond gauge.  In the boundary 
gauge, or any other gauges, appropriate gauge transformations must be applied 
to $\ket{\bj_0,\bk}$ before we can use this Bloch basis to block diagonalize 
the Hamiltonian.  Because of this, the computational cost for exact 
diagonalization incurred in the bond gauge is fractionally lower than in 
other gauges.

We can also calculate the correlation functions $\braket{\tilde{O}(\bR)
\tilde{O}'(\bR+\Delta\bR)}$ (of which the cluster DM is a function of) in 
any gauge, with appropriately-defined covariant observables $\tilde{O} = 
U O U^{\dagger}$, where $O$ are the `physical' observables we would use 
when there is no twist in the boundary conditions.  In the boundary gauge, 
these covariant observables $\tilde{O} = O$ and $\tilde{O}' = O'$ are 
particularly simple, except when the displacement vector $\Delta\bR$ crosses
the boundaries of our system.  For our purpose of calculating the cluster 
DM, this situation occurs only when the cluster itself straddle the system 
boundaries.  Therefore, with the cluster properly nested within the system,
we chose to perform twist boundary conditions averaging in the boundary gauge.

In the boundary gauge, the cluster DM is obtained by tracing
down the ground-state wave function $\ket{\Psi(\text{boundary})}$.  We can
get this wave function from $\ket{\Psi(\text{bond})}$ by applying the gauge 
transformation
\begin{equation}
\varphi: \ket{\bn} \to e^{-i\sum_{\bR} n_{\bR} \bR\cdot\bphi}\ket{\bn},
\end{equation}
where $\ket{\bn}$ is an occupation number basis state, with occupation
$n_{\bR}$ on site $\bR$.

Now, averaging over twist vectors $\bphi$ is the same as integrating over
the Brillouin Zone, so we perform twist boundary conditions averaging over 
a uniform grid of Monkhorst-Pack points with order $q$.\cite{monkhorst76}
For rectangular systems $(L_x, 0)\times(0, L_y)$, the First Brillouin Zone 
is sampled by varying the two independent twist angles between $-\pi/L_x 
\leq \phi_x < +\pi/L_x$ and $-\pi/L_y \leq \phi_y < +\pi/L_y$.  For non-%
square systems, the two independent twist angles $\phi_1$ and $\phi_2$ are
defined by
\begin{equation}
e^{i\bphi\cdot\bR_1} = e^{i\phi_1}, \quad
e^{i\bphi\cdot\bR_2} = e^{i\phi_2}.
\end{equation}
The twist vector $\bphi$ is then related to the independent twist angles 
$-\pi \leq \phi_1 < +\pi$ and $-\pi \leq \phi_2 < +\pi$ by
\begin{equation}
\bphi = \frac{\phi_1}{2\pi}\,\bQ_1 + \frac{\phi_2}{2\pi}\,\bQ_2,
\end{equation}
where
\begin{equation}
\bQ_1 = \frac{2\pi}{N}(R_{2y}, -R_{2x}), \quad
\bQ_2 = \frac{2\pi}{N}(-R_{1y}, R_{1x})
\end{equation}
are the primitive reciprocal lattice vectors of our non-square system.
For such systems, the First Brillouin Zone is a parallelogram on the
$\phi_x$-$\phi_y$ plane, so the uniform grid of Monkhorst-Pack points are 
imposed on the square domain $(-\pi,+\pi)\times(-\pi,+\pi)$ on
the $\phi_1$-$\phi_2$ plane instead.

\subsection{Classifying States of the Cluster}
\label{sect:states}

\begin{figure}[htbp]
\centering
\includegraphics{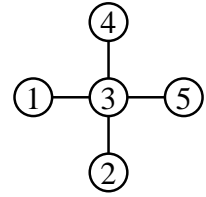}
\caption{The five-site, cross-shaped cluster whose cluster DM
we are interested in calculating for both a system of noninteracting, as
well as strongly-interacting spinless fermions.}
\label{fig:cross}
\end{figure}

The point symmetry group of the square lattice is the dihedral group
$D_4$, which has eight elements.\cite{dihedral}  The five
irreducible representations of this group are $A_1$, $A_2$, $B_1$, $B_2$
and $E$.  For the cross-shaped cluster shown in \figref{fig:cross}, there
is a one-to-one correspondence between these five irreducible representations
and the one-particle states, but instead of labeling these one-particle
states as $\ket{A_1}$, $\ket{A_2}$, $\ket{B_1}$, $\ket{B_2}$ and $\ket{E}$,
we adopt an angular momentum-like notation,
\begin{equation}
\begin{aligned}
\ket{s_+} &= \ket{A_1} = \tfrac{1}{\sqrt 5}(1, 1, 1, 1, 1), \\
\ket{s_-} &= \ket{A_2} = \tfrac{1}{\sqrt 5}(1, 1, -1, 1, 1), \\
\ket{p_x} &= \ket{B_1} = \tfrac{1}{\sqrt 2}(1, 0, 0, 0, -1), \\
\ket{p_y} &= \ket{B_2} = \tfrac{1}{\sqrt 2}(0, 1, 0, -1, 0), \\
\ket{d} &= \ket{E} = \tfrac{1}{2}(1, -1, 0, -1, 1).
\end{aligned}
\end{equation}
that would make clear the structure of these one-particle states.  We find
it more convenient to work with the one-particle states
\begin{equation}
\begin{aligned}
\ket{s} &= \tfrac{1}{\sqrt 2}\ket{s_+} + \tfrac{1}{\sqrt 2}\ket{s_-} =
\tfrac{1}{2}(1, 1, 0, 1, 1), \\
\ket{\bar s} &= (0, 0, 1, 0, 0), \\
\ket{p_+} &= \tfrac{1}{\sqrt 2}\ket{p_x} + \tfrac{i}{\sqrt 2}\ket{p_y} =
\tfrac{1}{2}(1, i, 0, -i, -1), \\
\ket{p_-} &= \tfrac{1}{\sqrt 2}\ket{p_x} - \tfrac{i}{\sqrt 2}\ket{p_y} =
\tfrac{1}{2}(1, -i, 0, i, -1).
\end{aligned}
\end{equation}

For a cluster DM possessing the full point group symmetry of
the square lattice, the one-particle state $\ket{d}$ is constrained by
symmetry to always be an eigenstate of $\rho_C$.  We call the associated
eigenvalue $w_d$ the \emph{weight} of $\ket{d}$.
Furthermore, the one-particle states $\ket{p_x}$ and $\ket{p_y}$ are also
equivalent under the square lattice symmetry, and hence their weights 
$w_{p_x}$ and $w_{p_y}$ are equal.  We call this doubly-degenerate 
one-particle weight $w_p$.  On the other hand, the $s$ one-particle
eigenstates of $\rho_C$ are in general not $\ket{s_+}$, $\ket{s_-}$ or 
$\ket{s}$, $\ket{\bar s}$, but some admixture of the form
\begin{equation}
\begin{aligned}
\ket{s_1} &= \cos\theta\ket{s} + \sin\theta\ket{\bar s}, \\
\ket{s_2} &= -\sin\theta\ket{s} + \cos\theta\ket{\bar s}.
\end{aligned}
\end{equation}
We call their corresponding weights $w_{s_1}$ and $w_{s_2}$ respectively.

We can then extend this angular-momentum-like notation to multi-particle
states of the cluster.  Though the quantum numbers used to label the
one-particle states are, strictly speaking, not angular momentum quantum
numbers, we apply the rules of angular momenta addition as if they were
to write down the angular-momentum-like quantum numbers for the multi-%
particle states.  For example, for the two-particle states of the cluster,
we have 
\begin{equation}\label{eqn:twoparticlestates}
\begin{gathered}
\ket{S} = \ket{p_+p_-}, \quad
\ket{S'} = \ket{s\bar s}, \\
\ket{P_+} = \ket{sp_+}, \quad
\ket{P'_+} = \ket{\bar sp_+}, \quad
\ket{P''_+} = \ket{p_-d}, \\
\ket{P_-} = \ket{sp_-}, \quad
\ket{P'_-} = \ket{\bar sp_-}, \quad
\ket{P''_-} = \ket{p_+d}, \\
\ket{D} = \ket{sd}, \quad
\ket{D'} = \ket{\bar sd}.
\end{gathered}
\end{equation}

\section{Noninteracting System}
\label{sect:ni}

In preparation for our main calculations on the strongly-interacting
system, we investigated in great details the cluster DM spectra of a 
system of noninteracting spinless fermions described by the Hamiltonian
\begin{equation}\label{eqn:nonintham}
H = -t\sum_{\braket{i,j}}\cxd{i}\cx{j}.
\end{equation}
An analytical formula for the cluster DM $\rho_C$ is known for this
system,\cite{cheong04a} using which we can obtain the spectrum of
$\rho_C$ for any system size.  We take advantage of this analytical 
formula to calculate $\rho_C$ for an infinite system of noninteracting 
spinless fermions.

However, our goal in calculating the cluster DM spectrum of 
noninteracting spinless fermions is to compare it against the cluster
DM spectrum of interacting spinless fermions.  For the latter
system, we can only calculate --- sans approximations --- the cluster 
DM from the exactly-diagonalized ground-state wave function of finite 
systems.  To make this comparison between noninteracting and 
interacting spinless fermions more meaningful, we compute their 
cluster DMs for the same series of finite systems.  In Section
\ref{sect:nicdm}, we describe how the infinite-system and finite-system 
cluster DMs for noninteracting spinless fermions are calculated.

In Section~\ref{sect:nifse}, we observe that the noninteracting
finite-system cluster DM spectra are contaminated by various
finite-size effects.  These finite-size effects also arise when we 
compute the interacting finite-system cluster DM spectra, so we want 
to learn how to deal with them.  Clearly, the effectiveness of various 
techniques in reducing finite-size effects can be best gauged by 
applying them to finite systems of noninteracting spinless fermions, 
since we will then be able to compare the results from the various
techniques against the infinite-system limit.  The simplest antidote 
to the various finite-size effects is to use a larger finite system.  
As expected, finite-size effects do become less and less important as 
the size of the system is increased.  Unfortunately, based on
comparisons of the finite-system cluster DM spectra with the
infinite-system cluster DM spectrum, we realized that we would need to 
go to system sizes of a few hundred sites in order for the finite-%
system cluster DM spectra to be decent approximations of the 
infinite-system cluster DM spectrum.  

Since such system sizes are not practical for exactly diagonalizing 
the strongly-interacting system given by \eqref{eqn:strongintham}, we 
look into the method of twist boundary conditions averaging in 
Section~\ref{sect:tbca}.  This method involves averaging the cluster 
DM spectra over various phase twists introduced into the periodic 
boundary conditions imposed on a given finite system.  For 
noninteracting spinless fermions, we find that twist boundary
conditions averaging reduces finite domain and shell effects in the 
cluster DM spectra, which then approximate the infinite-system cluster 
DM very well.  As a matter of standardization, we apply the method of 
twist boundary conditions averaging to both noninteracting and 
interacting spinless fermions, and compare their twist boundary 
conditions-averaged cluster DM spectra.

\subsection{Calculating the Noninteracting Cluster DM}
\label{sect:nicdm}

Instead of the general formalism presented in Section~\ref{sect:cdmf},
for the system of noninteracting spinless fermions we calculate the 
cluster DM weights using the exact formula
\begin{equation}\label{eqn:basicformula}
\rho_C = \det(\openone - G_C) \exp\biggl\{\sum_{i,j}
\log_e\left[G_C(\openone - G_C)\right]_{ij}
\cxd{i}\cx{j}\biggr\}
\end{equation}
obtained in Ref.\phantom{i}\onlinecite{cheong04a}, which relates the
DM $\rho_C$ of a cluster of sites and the cluster Green-function matrix 
$G_C$.  The matrix elements of $G_C$ are given by
\begin{equation}\label{eqn:twopointfunctionsum}
G_C(\bR, \bR') = \braket{\Psi|\cxd{\bR}\cx{\bR'}|\Psi} =
\frac{1}{N}\sum_{\text{$\bk$ filled}} e^{\im\bk\cdot(\bR - \bR')},
\end{equation}
where $\ket{\Psi}$ is the ground state of the system, and $\bR$, $\bR'$ 
are sites within the cluster.  The corollary of \eqref{eqn:basicformula} 
is that, if $\lambda_l$ is an eigenvalue of the cluster Green-function 
matrix $G_C$, the corresponding one-particle weight of $\rho_C$ is
\begin{equation}\label{eqn:wlll}
w_l = \lambda_l \prod_{l'\neq l} (1 - \lambda_{l'}).
\end{equation}

To calculate the infinite-system spectra of $\rho_C$, we convert the
sum over $\bk_n$ in \eqref{eqn:twopointfunctionsum} into an integral 
\begin{equation}\label{eqn:infsystwopointfunction}
G_C(\bR - \bR') = \int_{\epsilon(\bk) \leq \epsilon_F}
\frac{d^2\bk}{(2\pi)^2}\, e^{\im\bk\cdot(\bR - \bR')}
\end{equation}
over those wave vectors $\bk$ bounded by the Fermi surface
$\epsilon(\bk) = \epsilon_F$, where $\epsilon_F$ is the Fermi energy.
On an infinite square lattice with unit lattice constant, the
dispersion relation is given by
\begin{equation}
\epsilon(\bk) = -2\left(\cos k_x + \cos k_y\right).
\end{equation}
We then obtain the infinite-system cluster Green-function matrix 
eigenvalues $\lambda_{s_1}(\epsilon_F)$, $\lambda_p(\epsilon_F)$,
$\lambda_d(\epsilon_F)$ and $\lambda_{s_2}(\epsilon_F)$ as 
functions of the Fermi energy $\epsilon_F$ by numerically integrating
\eqref{eqn:infsystwopointfunction}, and diagonalizing the resulting
infinite-system cluster Green-function matrix $G_C(\epsilon_F)$.  For 
the same set of Fermi energies, we also integrate
\begin{equation}
\nbar(\epsilon_F) = \int_{\epsilon(\bk) \leq \epsilon_F}
\frac{d^2\bk}{(2\pi)^2},
\end{equation}
over the Fermi surfaces to find the corresponding filling fractions.
The one-particle infinite-system cluster DM weights
$w_{s_1}(\epsilon_F)$, $w_p(\epsilon_F)$, $w_d(\epsilon_F)$, and
$w_{s_2}(\epsilon_F)$ are then calculated using \eqref{eqn:wlll}.

To calculate the cluster DM spectra for a finite system of $N$ sites
with $P$ noninteracting particles, we determine the set of wave 
vectors $\{\bk_n\}_{n=1}^P$ with the lowest single-particle energies
\begin{equation}
\epsilon_{\bk_n} = -2(\cos k_{n,x} + \cos k_{n,y}).
\end{equation}
We then evaluate the finite-system cluster Green-function matrix 
elements in \eqref{eqn:twopointfunctionsum} by summing over these 
occupied wave vectors $\{\bk_n\}_{n=1}^P$.  Following this, we
diagonalize the finite-system cluster Green-function matrix $G_C(P)$ 
to obtain the eigenvalues $\lambda_{s_1}(P)$, $\lambda_p(P)$, 
$\lambda_d(P)$, and $\lambda_{s_2}(P)$, and therefrom the one-particle
finite-system cluster DM weights $w_{s_1}(P)$, $w_p(P)$, $w_d(P)$, and
$w_{s_2}(P)$ using \eqref{eqn:wlll}.  By varying the number $P$ of 
noninteracting particles in the finite $N$-site system, we determined
the finite-system cluster DM one-particle spectra for the filling
fractions $\nbar = P/N$ accessible to the finite system.

\subsection{Finite Size Effects and the Infinite-System Limit}
\label{sect:nifse}

\begin{figure}[htbp]
\centering
\includegraphics[scale=0.4,clip=true]{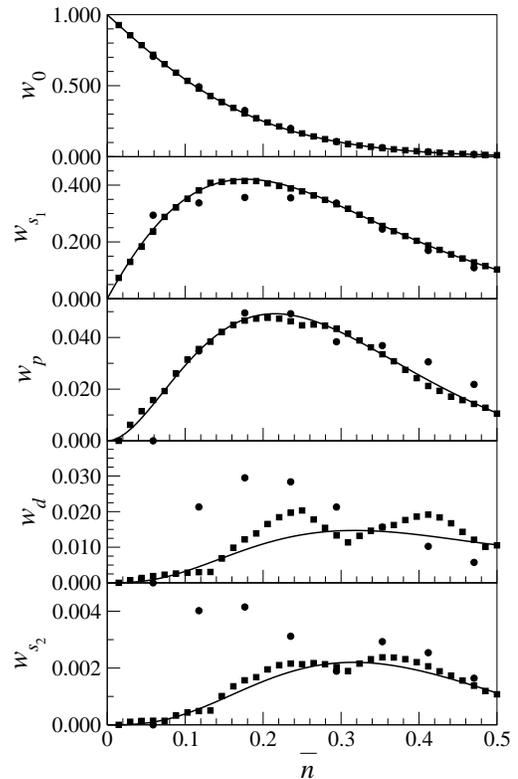}
\caption{Zero- and one-particle weights of the cluster DM of a
5-site, cross-shaped cluster for systems of noninteracting spinless 
fermions with periodic boundary conditions imposed.  The finite 
$(4,-1)\times(1,4)$ (\CIRCLE) and $(8,-2)\times(2,8)$ ($\blacksquare$) 
systems have the same shape but different sizes.  The solid curves are
the zero- and one-particle weights for the infinite system.  For 
$w_{s_1}$, we see the finite domain effect deviations for the $(4,-1)
\times(1,4)$ system is practically gone by the time we get to the 
$(8,-2)\times(2,8)$ system.  For the rest of the one-particle weights, 
the finite domain effect is largely removed in the $(8,-2)\times(2,8)$ 
system, but shell effect persists.  In fact, shell effect in $w_d$ is 
still visible when we go to the $(32,-8)\times(8,32)$ system (not 
shown), which has $N = 1088$ sites.}
\label{fig:free0n1}
\end{figure}

Imposing the usual periodic boundary conditions, we calculated the
finite-system spectra of $\rho_C$ for several small systems ranging 
from $N = 13$ to $N = 20$ sites.  For these small systems sizes, we 
find that it is impossible to say anything meaningful about the
dependence on the filling fraction $\nbar = P/N$ for the cluster DM
weights because of the finite size effects.  Using the relation 
\eqref{eqn:basicformula} for noninteracting systems, we investigated the 
effect of system size on the spectrum of $\rho_C$ for a series of system 
$(4p, -p)\times(p, 4p)$, $1 \leq p \leq 8$, with the same shape.  As the 
system size is increased, we find that the cluster DM spectrum 
approaches the infinite-system limit, as shown in Figure 
\ref{fig:free0n1}.

For this series of systems, the infinite-system limit is more or less 
reached at around $p = 4$ (272 sites), based on comparison with the
infinite-system limit itself.  We can also arrive at this estimate by
looking at the convergence of the one-particle weights alone.  More
importantly, we find that the shell effect affects weights of 
different symmetry differently: $w_{s_1}$ is almost unaffected, while 
$w_d$ is the most severely affected.  Shell effect persists in $w_d$ 
even up to a system size of 1088 sites (for $p = 8$).

We also looked at $w_{s_1}(\nbar)$, which is almost unaffected by shell
effect, for several systems with between 200 to 300 sites of different 
shapes.  For systems of these sizes, the finite domain effect is negligible,
but we find $w_{s_1}(\nbar)$ from finite systems of different shapes
differing very slightly from the infinite-system limit, and also from
each other.  Since we expect systems of different shapes to all 
approach the same infinite-system limit, we attribute these very small 
deviations to the shape effect.  Based on more extensive numerical
studies (see Chapter 4 of Ref. \onlinecite{cheong05}) not reported in 
this paper, we know that shape effect deviations are not effectively 
removed by the three averaging devices we have introduced in 
Section~\ref{sect:averaging}, but fortunately these deviations are
very small.

\subsection{Twist Boundary Conditions Averaging}
\label{sect:tbca}

For a system of interacting spinless fermions, we cannot directly 
compute the exact infinite-system cluster DM.  We can, however,
choose to work with (i) an approximate ground-state wave function
of an infinite system, or (ii) the exact ground-state wave function of
a finite system, or (iii) an approximate ground-state wave function of 
a finite system.  As reported in Section \ref{sect:strong}, we chose 
option (ii), where the exact ground-state wave function is obtained 
through numerical exact diagonalization.

Exact diagonalization severely limits the sizes of the finite systems 
we can work with (see Appendix of Ref.~\onlinecite{zhang03} for
formula on size of Hilbert space for the strongly-interacting model
given by \eqref{eqn:strongintham}).  With aggressive memory reduction 
measures, it is possible (but not necessarily feasible) to exactly 
diagonlize finite systems with up to 30 sites for all filling 
fractions.  However, as we have illustrated in 
Section \ref{sect:nifse}, for systems so small, the numerical cluster 
DM spectra would be plagued by strong finite size effects, most 
notably by the shell effect.  This is where twist boundary conditions 
averaging comes in.

Very crudely speaking, we can think of averaging numerical observables
over $M$ twist boundary conditions for a finite system with $N$ sites 
as being equivalent to computing these numerical observables for a 
single finite system of $N^* > N$ sites, subject to only periodic 
boundary conditions.  In the best-case scenario, the effective system
size $N^*$ might be as large as $MN$, though $N^*$ will typically grow
slower than $O(M)$.  The computational cost of performing exact
diagonalization for a system with $N$ sites over $M$ twist boundary 
conditions is on the order of $O(MN^3)$, whereas the computational
cost of performing exact diagonalization just once for a system with
$N^* > N$ sites is $O({N^*}^3)$.  So long as the effective system size 
$N^*$ grows faster than $O(M^{1/3})$, it would be computational
cheaper to employ the method of twist boundary conditions averaging,
instead of exactly diagonalizing a single large finite system, to
reduce finite size effects.  The detail dependence of $N^*$ on $M$ 
will of course depend on the nature of the observable of interest.

From the detailed study undertaken in Appendix D of Ref.
\onlinecite{cheong05}, we know that there are cuts and cusps on the twist 
surface $\braket{\Psi(\phi_x,\phi_y)|O|\Psi(\phi_x, \phi_y)}$ of a 
generic observable $O$, where $\ket{\Psi(\phi_x, \phi_y)}$ is the 
many-body ground state of a finite $N$-site system subject to twist 
boundary conditions with twist vector $\bphi = (\phi_x, \phi_y)$.  For
non-square systems, these cusps and cuts demarcate features with a hierarchy 
of sizes on the twist surface.  The `typical' twist surface feature has a
linear dimension of $2\pi/\sqrt{N}$.  These are decorated by fine structures
with linear dimension $2\pi/N$, which are in turn decorated by hyperfine 
structures with linear dimensions $2\pi/N^2$.  The number of integration
points we must use is therefore determined by what feature size we
want to integrate faithfully.

\begin{figure}[htbp]
\centering
\includegraphics[scale=0.4,clip=true]{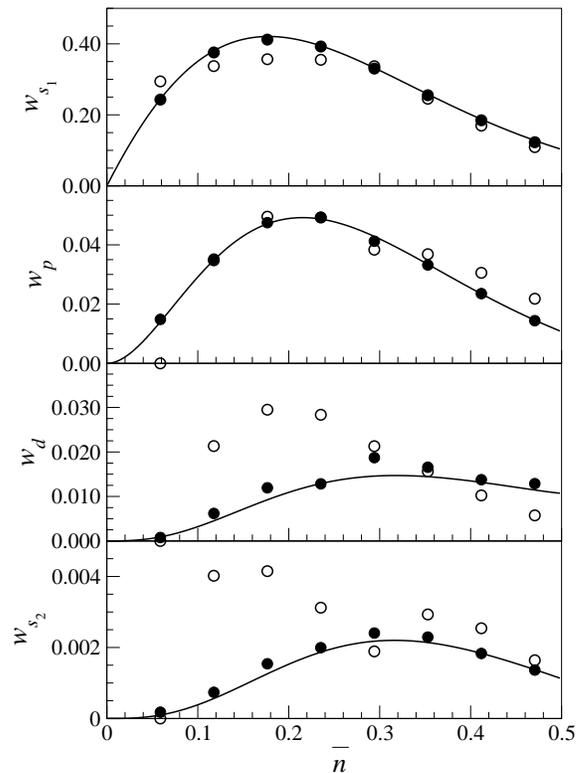}
\caption{One-particle weights of the cluster DM of a 5-site, 
cross-shaped cluster within systems of noninteracting spinless 
fermions.  The performance of twist boundary conditions averaging, 
using $q = 16$ Monkhorst-Pack special-point integration, in reducing 
finite size effects for the $(4,-1)\times(1,4)$ system (\CIRCLE) is 
checked against the $(4,-1)\times(1,4)$ (\Circle) system with periodic 
boundary conditions imposed.  The solid curves are the cluster DM
weights from the infinite system.}
\label{fig:free1tbcav}
\end{figure}

\begin{figure}[htbp]
\centering
\includegraphics[scale=0.4,clip=true]{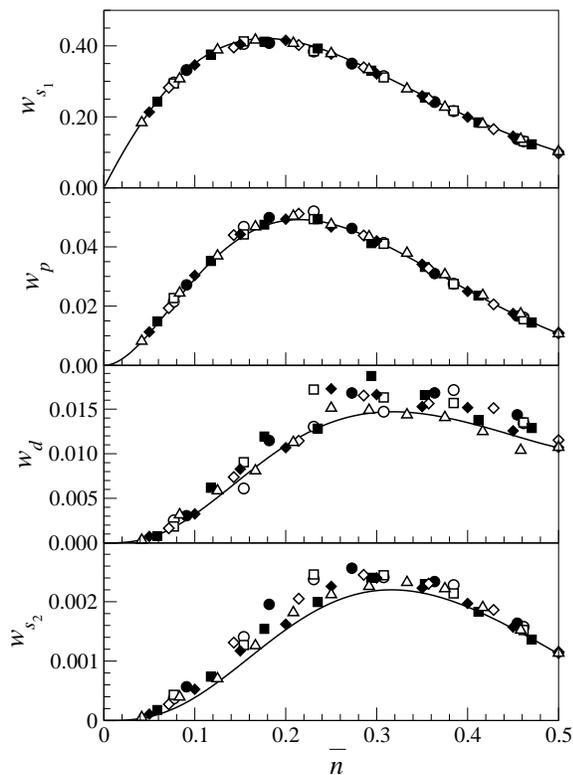}
\caption{One-particle weights of the cluster DM of a 5-site,
cross-shaped cluster, for the $(3,-2)\times(2,3)$ (\CIRCLE),
$(4,1)\times(1,3)$ (\Circle), $(4,-1)\times(1,3)$ ($\square$),
$(4,-1)\times(1,4)$ ($\blacksquare$), $(4,-1)\times(2,3)$
($\lozenge$), $(4,-2)\times(2,4)$ ($\blacklozenge$) and
$(5,1)\times(1,5)$ ($\bigtriangleup$) systems of noninteracting
spinless fermions subject to twist boundary conditions averaging,
using $q = 16$ Monkhorst-Pack special-point integration.  Also shown
are the cluster DM weights of the infinite system.}
\label{fig:free1diffsys} 
\end{figure}

For the purpose of this study, we decided to integrate the fine structure
on the twist surface faithfully.  Therefore, we chose to average the
spectrum of $\rho_C$ over a $q = 16$ Monkhorst-Pack grid (which consists
of 256 integration points in the First Brillouin Zone) for the $(4,-1)
\times(1,4)$ system with $N = 17$ sites.  We find that twist boundary 
conditions averaging does indeed result in an averaged spectrum which 
approximates the infinite-system limit well (see 
\figref{fig:free1tbcav}).  This averaging device, however, does not
completely eliminate shell effects, as can be seen from the twist boundary
conditions-averaged $w_d(\nbar)$.  To reduce the bias this creates for 
one particular choice of finite system, we combined the twist boundary
conditions-averaged spectrum of $\rho_C$ for
various finite systems.  This is shown in \figref{fig:free1diffsys}.
We will overlay the cluster DM spectra from several finite systems 
in the same way, to derive a twist boundary conditions-averaged 
approximation to the infinite-system cluster DM spectrum of
strongly-interacting spinless fermions.

\section{Strongly-Interacting System}
\label{sect:strong}

In this section, we compute the cluster DM for interacting spinless
fermions.  As with the case of noninteracting spinless fermions, the
cluster DM evaluated directly from the ED of various finite systems
are severely affected by finite size effects.  In Section
\ref{sect:ni}, we saw how finite size effects can be significantly
reduced when the cluster DM is averaged over various twist boundary
conditions, in the sense that the averaged cluster DM weights from
different finite systems at various filling fractions fall close to
their respective infinite-system limits.  Applying twist boundary
conditions averaging onto the interacting cluster DM, we find that
finite size effects are reduced, but not as dramatically as for the
noninteracting cluster DM.  Nevertheless, the averaged cluster DM
weights from different finite systems with different number of
particles are sufficiently consistent with each other that we can plot
a smooth curve interpolating the averaged cluster DM weights.

As explained in Section~\ref{sect:intro}, our interest in studying the
strongly-interacting model \eqref{eqn:strongintham} of spinless 
fermions with infinite nearest-neighbor repulsion is to understand how
the cluster DM evolves with filling, given that we expect in this
model crossovers between regimes of qualitatively different states.
Furthermore, we had proposed an operator-based method of truncation
which was justified by the fact that the noninteracting cluster DM is
generated from a set of single-particle operators.  Since we proposed
the use of this truncation scheme for interacting systems, we are
interested to know whether the structure of the interacting cluster DM
is such that it can also be generated, perhaps approximately, from a
set of single-particle operators.

To this end, we present in Sections~\ref{sect:strongzero},
\ref{sect:strongone} and \ref{sect:strongtwo}, the zero-, one- and
two-particle cluster DM weights of the strongly-interacting system of
spinless fermions.  We check whether it is possible to: (i) write
the two-particle eigenstates as the product of one-particle
eigenstates; and (ii) predict the relative ordering of the
two-particle weights based on the relative ordering of the
one-particle weights.  We then discuss in Section \ref{sect:FLDM}
whether these two criteria are met, and whether it is feasible to
design an operator-based DM truncation scheme, similar to the one
described in Ref.~\onlinecite{cheong04b}, for the strongly-interacting
system.

\subsection{Zero-Particle Cluster DM Weight}
\label{sect:strongzero}

\begin{figure}[htbp]
\centering
\includegraphics[scale=0.4,clip=true]{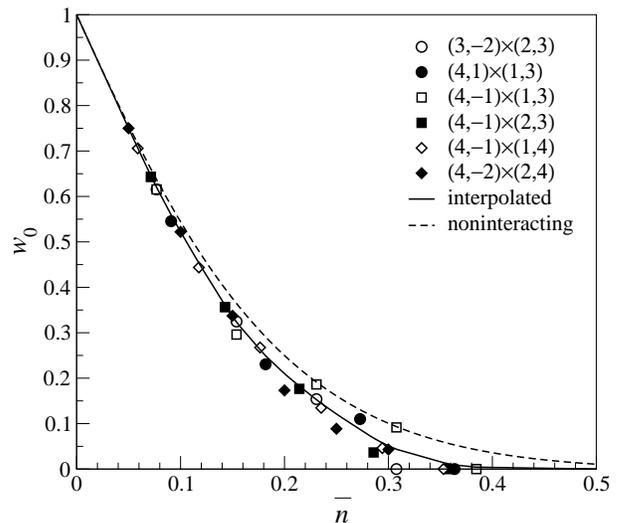}
\caption{Zero-particle weight of the cluster DM of a 5-site, cross-%
shaped cluster, for the $(3,-2)\times(2,3)$ (\CIRCLE), $(4,1)\times
(1,3)$ (\Circle), $(4,-1)\times(1,3)$ ($\square$), $(4,-1)\times(1,4)$
($\blacksquare$), $(4,-1)\times(2,3)$ ($\lozenge$) and $(4,-2)\times
(2,4)$ ($\blacklozenge$) systems of strongly-interacting spinless 
fermions subject to twist boundary conditions averaging, using $q = 8$ 
Monkhorst-Pack special-point integration.  At $\nbar = 0$ and $\nbar = 
\frac{1}{2}$, we know analytically that $w_0 = 1$ and $w_0 = 0$ 
respectively, and the solid `curve' interpolates between these two 
known limits and the equally weighted data points at finite filling 
fractions $0 < \nbar < \frac{1}{2}$.  Also shown as the dashed curve 
is the zero-particle weight of the infinite system of noninteracting 
spinless fermions.}
\label{fig:w0strong}
\end{figure}

The zero-particle cluster DM weight calculated for various 
finite strongly-interacting systems is shown in \figref{fig:w0strong}.  
Also shown in the figure is the zero-particle cluster DM weight of the 
infinite system of noninteracting spinless fermions.  As we can
see, the zero-particle weights of the respective systems only start 
differing significantly from each other for $\nbar > 0.1$.  With repulsive 
interacting between spinless fermions, it is more difficult in a congested 
system ($\nbar > 0.2$) to form an empty cluster of sites from quantum 
fluctuations.  As a result, the strongly-interacting $w_0$ falls below 
the noninteracting $w_0$.  However, this fact alone does not tell us 
anything more about the correlations in the strongly-interacting ground 
state, and so we move on to consider the one-particle cluster DM weights.

\subsection{One-Particle Cluster DM Eigenstates and Their Weights}
\label{sect:strongone}

In our Operator-Based DM Truncation Scheme developed in
Ref.~\onlinecite{cheong04b} for a noninteracting system, the
one-particle cluster DM weights play a very important role, since we
select which one-particle operators to keep or discard based on the
negative logarithm of these numbers.  We expect the one-particle
cluster DM weights, though not entirely sufficient by themselves,
would also play an important role in defining an operator-based
truncation scheme.  Therefore, in this section, we present results for
a series of calculations to determine the infinite-system limit of the
one-particle cluster DM spectra for our strongly-interacting system as
a function of filling fraction $\nbar$, and discuss their implications
for an operator-based DM truncation scheme.

\begin{figure}[htbp]
\centering
\includegraphics[scale=0.4,clip=true]{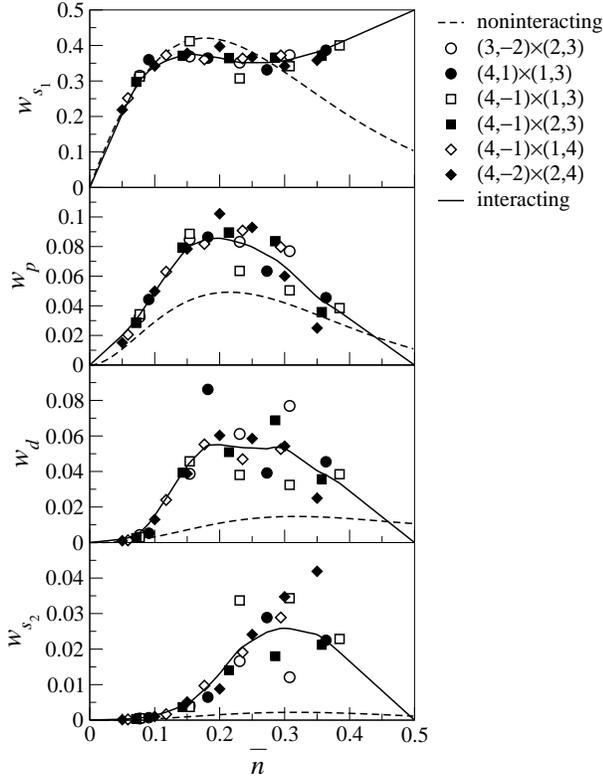}
\caption{One-particle weights of the cluster DM of a 5-site, 
cross-shaped cluster within a system of strongly-interacting spinless 
fermions.}
\label{fig:w1stronginterpolate}
\end{figure}

Though we really do need to worry about the evolution of the structure
of $\ket{s_1}$ and $\ket{s_2}$ as a function of $\nbar$ in both the
noninteracting and strongly-interacting systems, the one-particle weights
are ordered by their magnitudes as $w_{s_1} > w_p > w_d > w_{s_2}$ for both 
systems.  But while the noninteracting one-particle weights go down by 
roughly one order of magnitude as we go through the sequence $w_{s_1} \to 
w_p \to w_d \to w_{s_2}$, we see from \figref{fig:w1stronginterpolate} that
the interacting one-particle weights decay more slowly along this same 
sequence.

We studied the finite $(3,-2)\times(2,3)$ (\CIRCLE), $(4,1)\times(1,3)$ 
(\Circle), $(4,-1)\times(1,3)$ ($\square$), $(4,-1)\times(1,4)$ 
($\blacksquare$), $(4,-1)\times(2,3)$ ($\lozenge$) and $(4,-2)\times(2,4)$ 
($\blacklozenge$) systems subject to twist boundary conditions averaging, 
using $q = 8$ Monkhorst-Pack special-point integration.  At a filling 
fraction of $\nbar = 0$, the system approaches the noninteracting limit, 
and thus all the one-particle weights are zero.  At half-filling $\nbar = 
\frac{1}{2}$, the two-fold degenerate checker-board ground state is 
unaffected by twist boundary conditions averaging.  We can thus perform 
degeneracy averaging analytically, to find that $w_{s_1} = \frac{1}{2}$ 
and $w_p = w_d = w_{s_2} = 0$.  The solid `curves' in Figure
\ref{fig:w1stronginterpolate} interpolate between these two known limits 
and the equally weighted data points at finite filling fractions $0 < 
\nbar < \frac{1}{2}$.  Also shown in Figure \ref{fig:w1stronginterpolate}
as the dashed curves are the one-particle weights of the infinite
system of noninteracting spinless fermions.

When only periodic boundary conditions are imposed, there is 
significantly more `scatter' in the one-particle weights as a function 
of filling fraction $\nbar$, for interacting systems of different 
sizes, than for noninteracting systems of different sizes.  Averaging 
the one-particle weights of the interacting systems over various twist 
boundary conditions visibly reduces this `scatter', even though the 
remnant `scatter' seen in Figure \ref{fig:w1stronginterpolate} is 
still rather large, compared to case for the noninteracting system
(Figure \ref{fig:free1diffsys}).  From our own detailed study of the 
method of twist boundary conditions averaging for noninteracting
systems,\cite{cheong05} we know that twist boundary conditions 
averaging effectively removes finite size effects from some 
observables, but not for others.  We have no reason to expect twist 
boundary conditions averaging to apply equally effectively over the
same set of observables, when we go from noninteracting systems to
interacting systems.  Conversely, observables for which twist boundary
conditions averaging is ineffective in noninteracting systems, might
be effectively twist-boundary-conditions-averaged in interacting
systems.  With only input from the exact diagonalization of finite 
systems, and without employing system-size extrapolations, the method 
of twist boundary conditions averaging offers us the best hope of
gaining insight into the infinite-system properties we seek.

We expect that the remnant `scatter' in the twist-boundary-%
conditions-averaged one-particle weights will be reduced, if we had
not made the nearest-neighbor repulsion infinite.   There are two
reasons why we did not also study the case of $t \ll V < \infty$.
First, for a fixed system size $N$ and particle number $P$, the 
Hilbert space for the $V < \infty$ system would be much larger than 
that of the $V \to \infty$ system.  A parallel study for $V < \infty$
of the $V \to \infty$ system sizes and particle numbers reported in 
this paper, with twist boundary conditions averaging, will require 
unacceptably long total computation time.  Second, and more 
importantly, we believe that as long as $V/t < \infty$ is large, the
qualitative implications on the Operator-Based DM Truncation Scheme
would be similar to the case where $V/t \to \infty$, and therefore it
suffices to examine the latter case, which is computationally much
more manageable.  In any case, we do not believe the remnant `scatter'
in the twist-boundary-conditions-averaged DM weights will hamper our
efforts in drawing qualitative conclusions regarding the
applicability, or otherwise, of the Operator-Based DM Truncation
Scheme for interacting Fermi liquids.

In the Operator-Based DM Truncation Scheme described in
Ref.~\onlinecite{cheong04b}, we discard one-particle cluster 
DM eigenstates with very small weights, and keep only the many-particle 
cluster DM eigenstates built from the retained one-particle eigenstates.  
The sum of weights of the truncated set of cluster DM eigenstates will 
then be very nearly one, \emph{if} the discarded one-particle weights are
all very small compared to the maximum one-particle weight.  As we can
see from \figref{fig:w1stronginterpolate}, the ratio of the largest
one-particle weight, $w_{s_1}$, to the smallest one-particle weight,
$w_{s_2}$, is not large enough for us to justify keeping $\ket{s_1}$
and discarding $\ket{s_2}$, except when the system is very close to
half-filled.

We believe that the one-particle cluster DM weights are so close to
each other in magnitude, because of the net `strength' of interactions
straddling the cluster and its environment is strong compared to the
net `strength' of interactions strictly within the cluster.
Unfortunately for our five-site cluster, which was chosen because it
is the smallest non-trivial cluster having the full point group
symmetry of the underlying square lattice,\footnote{The $2 \times 2$
cluster is the smallest cluster having $D_4$ as its point symmetry
group.  However, it is trivial as a cluster, because we can have at
most $P = 2$ particles within this cluster, as opposed to a maximum 
of $P = 4$ particles within the five-site cluster.} the sites within
the cluster are poorly connected, i.e. a cluster site is on average
connected to more environmental sites than to other cluster sites.
To have more of the total interactions of the cluster with the system
be confined within the cluster, a cluster significantly larger than 
the five-site cluster studied in this paper will be needed.  This
large cluster must then be embedded in a finite system that is larger 
still, making exact diagonalization studies unfeasible.

\subsection{Two-Particle Cluster DM Eigenstates and Their Weights}
\label{sect:strongtwo}

\begin{figure}[htbp]
\centering
\includegraphics[scale=0.4,clip=true]{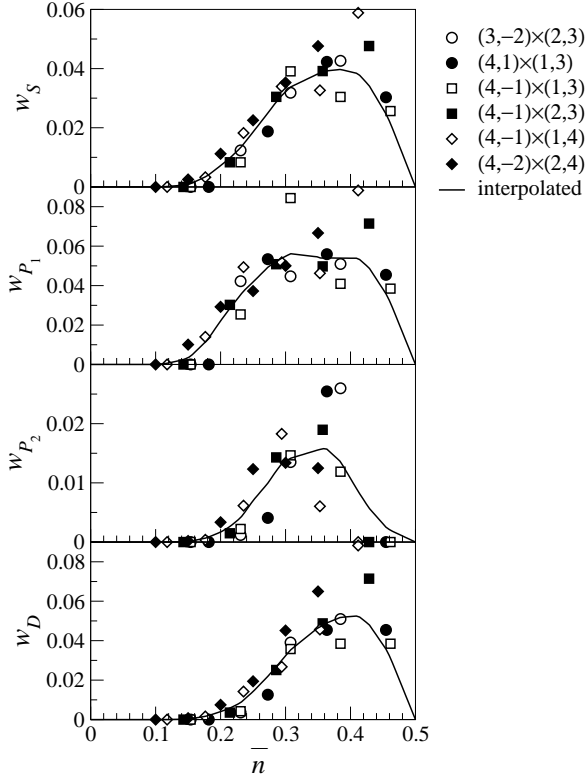}
\caption{Two-particle weights of the cluster DM of a 5-site, 
cross-shaped cluster within a system of strongly-interacting spinless 
fermions.}
\label{fig:w2strong}
\end{figure}

While it is desirable to have a broader distribution of one-particle 
weights, our more important task is to examine how closely the
many-particle cluster DM eigenstates can be approximated as products of
one-particle cluster DM eigenstates.  In particular, we look at the
two-particle cluster DM eigenstates, and find that of the two-particle 
states listed in \eqref{eqn:twoparticlestates}, the only states which 
are allowed by the no-nearest-neighbor constraint to appear in the 
cluster Hilbert space are $\ket{S}$, $\ket{P_{\pm}}$, $\ket{P''_{\pm}}$ 
and $\ket{D}$.  We know therefore that the two-particle sector of 
$\rho_C$ comprises a $1\times 1$ $S$-diagonal block (with weight
$w_S$), a $1\times 1$ $D$-diagonal block (with weight $w_D$), and two 
degenerate $2\times 2$ $P$-diagonal blocks (with weights $w_{P_1}$ and
$w_{P_2}$).  The two-particle weights are shown as a function of filling
$\nbar$ in \figref{fig:w2strong}.

For the finite $(3,-2)\times(2,3)$ (\CIRCLE), $(4,1)\times(1,3)$ 
(\Circle), $(4,-1)\times(1,3)$ ($\square$), $(4,-1)\times(1,4)$ 
($\blacksquare$), $(4,-1)\times(2,3)$ ($\lozenge$) and $(4,-2)\times(2,4)$ 
($\blacklozenge$) systems studied, subject to twist boundary conditions 
averaging, using $q = 8$ Monkhorst-Pack special-point integration, all
the two-particle weights are zero at $\nbar = 0$ as the systems approach
the noninteracting limit.  At half-filling $\nbar = \frac{1}{2}$, 
we again perform degeneracy averaging analytically on the two-fold 
degenerate checker-board ground state to find that all the two-particle 
weights are zero.  In Figure \ref{fig:w2strong}, the solid `curves' 
interpolates between these two known limits and the equally weighted data 
points at finite filling fractions $0 < \nbar < \frac{1}{2}$.

We realized that there are significantly fewer nontrivial two-particle 
eigenstates of $\rho_C$ than predicted from the combination of one-%
particle eigenstates.  From the point of view of implementing the 
Operator-Based DM Truncation Scheme, this poses no problem \emph{if} 
the non-occurring two-particle states are predicted to have small 
enough weights that they will be excluded by the truncation scheme.
However, we find that this is not the case.  For example, the two-particle 
state $\ket{S'}$, which does not occur, is predicted by simple combination
of the one-particle states $\ket{s}$ and $\ket{\bar{s}}$ to have a weight
comparable to that of the two-particle state $\ket{S}$, which does occur.

Of the two-particle weights that are allowed by the no-nearest-neighbor 
constraint, we expect their weights to follow the sequence $w_{P_1} \gtrsim 
w_D > w_S$, if they can indeed to thought of products of one-particle states.
From \figref{fig:w2strong}, we indeed observe this sequence of two-particle
weights, even though their actual magnitudes (calculated as the product of 
one-particle weights divided by the zero-particle weight) do not come out 
right.  This observation is encouraging, because we might yet be able 
to push a variant of the Operator-Based DM Truncation Scheme through,
by introducing constraints on how many-particle cluster DM states can
be built up from one-particle cluster DM states.

\subsection{Signatures of Fermi-Liquid Behaviour in the Cluster DM}
\label{sect:FLDM}

Over broad ranges of filling fractions, the ground state of our
strongly-interacting model \eqref{eqn:strongintham} of spinless
fermions is expected to be an interacting Fermi liquid.  While we
understand the cluster DM structure of a noninteracting Fermi liquid
completely, we do not yet understand how an interacting Fermi liquid
will manifest itself in the structure of its cluster DM.  Unlike a
noninteracting Fermi liquid, the ground-state DM of an interacting
Fermi liquid will not simply be the exponential of a noninteracting
pseudo-Hamiltonian.  Nevertheless, we expect that the interacting
pseudo-Hamiltonian $\tilde{H}$ appearing in the interacting
Fermi-liquid ground-state DM $\rho = \exp(\tilde{H})$ can be made to
look like the sum of a noninteracting pseudo-Hamiltonian
$\tilde{H}_0$, and a much weaker interaction term $\tilde{H}_1$, by a
canonical transformation.  From Landau's Fermi liquid theory, we know
that such a canonical transformation (similar to the one which relates
Landau quasi-particles to bare fermions) works by burying much of the
bare interactions within the quasi-particles.

In tracing down the ground-state DM, our hope then is that the cluster 
DM can also be written, after a canonical transformation local to the
cluster, as the exponential of the sum of a noninteracting pseudo-%
Hamiltonian $\tilde{H}_{C, 0}$ (which is perhaps related to 
$\tilde{H}_0$ in the same way as for the noninteracting Fermi liquid),
and a weak interaction term $\tilde{H}_{C, 1}$.  We suspect that the
criterion for this to be possible is that we must be able to construct
approximate quasi-particles using \emph{only} cluster states to absorb
the bare interactions.  However, we also believe that the requirement
that the canonical transformation act strictly within the cluster will 
fail to completely incorporate interactions that straddle the cluster 
and its environment.

In Section \ref{sect:strongtwo}, we found that for our
strongly-interacting system, the two-particle cluster DM eigenstates
look nothing like simple products of two one-particle cluster DM
eigenstates each.  In fact, many combinations of two one-particle
cluster DM eigenstates give invalid two-particle cluster states that
violate the condition of no-nearest-neighbor occupation.  It is
tempting, based on this observation, to say then that the cluster DM
is not the exponential of an approximately noninteracting
pseudo-Hamiltonian.  However, it must be remembered that in an
interacting Fermi liquid, the quasi-particles are also not single bare
particles.  Instead, they are superpositions of states containing
different number of bare particles, which leads us to think of a
quasi-particle as a bare particle being screened by other bare
particles in its vicinity.

With this in mind, we realized that to identify the quasi-particle 
structure of the cluster DM, we need to construct appropriate linear 
combinations of the $P$-particle cluster DM eigenstates, so that the 
cluster DM, when written in terms of these `quasi-particles', look
like the exponential of a noninteracting pseudo-Hamiltonian 
$\tilde{H}_{C,0}$ plus a weak interaction term $\tilde{H}_{C,1}$.
This involves writing the pseudo-Hamiltonian $\tilde{H}_C$ as a sum of 
terms, representing the independent quantum fluctuations associated 
with each of the quasiparticles.  This can be accomplished by defining
an operator singular value decomposition of the cluster DM with
respect to an appropriate operator norm, which forms the basis for
judging whether the quantum fluctuations associated with two linear 
combination of bare operators are independent.  Details of such an
operator singular value decomposition will be reported in a future
paper.\cite{cheong07}

\section{Summary \& Discussions}
\label{sect:summary}

To summarize, we have calculated numerically the cluster DM for a 
cross-shaped cluster of five sites within both a system of
noninteracting spinless fermions described by the Hamiltonian
\eqref{eqn:nonintham}, and a system of strongly-interacting spinless
fermions described by the Hamiltonian \eqref{eqn:strongintham}.  For
the noninteracting system, the cluster DM was obtained from the 
cluster Green-function matrix using the exact formula
\eqref{eqn:basicformula} obtained in Ref.~\onlinecite{cheong04a},
whereas for the interacting system, the cluster DM was obtained from 
the exact-diagonalization ground-state wave function by tracing down 
degrees of freedom outside of the cluster.  For the purpose of making
the comparison of the cluster DM spectra more straightforward, we
worked with the same collection of finite non-square systems for
interacting and noninteracting spinless fermions.

To make the results of this comparison less dependent on the geometry 
of the finite systems chosen, degeneracy averaging followed by 
orientation averaging of the cluster DM spectra were carried out.  
When combined, these two averaging apparatus has the effect of 
restoring full square-lattice symmetry to the cluster DM.  We also 
analyzed in detail the finite size effects not removed by degeneracy 
and orientation averaging, by inspecting the numerical cluster DM
spectra coming from finite noninteracting systems of various sizes.
By increasing the system size systematically, we find visually that 
the infinite-system limit of the cluster DM is `attained' when the 
system size reaches a couple of hundred sites.  

Noting that these are forbiddingly large system sizes to work with for
interacting systems, where the ground-state wave functions have to be 
obtained via exact diagonalization, we then tested the apparatus of 
twist boundary conditions averaging on finite systems with between 10 
and 20 sites.  For noninteracting spinless fermions, we find that the 
twist boundary conditions-averaged cluster DM weights for different
finite systems and different filling fractions indeed fall close to 
the various infinite-system limits.  Since we do not perform
system-size extrapolations, we interpolate between the degeneracy-, 
orientation-, and twist boundary conditions-averaged cluster DM 
weights for the various finite interacting systems and their
respective accessible filling fractions, and take the result curves to
be our best approximation of the cluster DM spectrum of the infinite
interacting system.

Comparing the twist boundary conditions-averaged cluster DM spectra 
for the noninteracting and strongly-interacting systems, we find 
similar qualitative behavior in the zero-particle weights as functions
of filling fraction, and qualitatively different behaviours in the
one-particle weights as functions of filling fraction.  However, the
relative ordering $w_{s_1} > w_p > w_d > w_{s_2}$ is the same at all
$\nbar < \frac{1}{2}$ for both systems.  Quantitatively, we find for
noninteracting spinless fermions that the one-particle weights go down 
by roughly one order of magnitude each time as we go through the 
sequence $w_{s_1} \to w_p \to w_d \to w_{s_2}$.  For strongly-%
interacting spinless fermions, the one-particle weights decay much
more slowly along the sequence.

The implications this observation have for the Operator-Base DM
Truncation Scheme developed in Ref.~\onlinecite{cheong04b} is that,
for a small fixed fraction of one-particle eigenstates retained, the
total cluster DM weight of eigenstates retained would be much smaller
for the strongly-interacting system compared to the noninteracting
system, since the ratio of the smallest to the largest one-particle
weights, $w_{s_2}/w_{s_1}$, is not very much smaller than one.  This
narrow distribution of one-particle cluster DM weights aside, we
observed that the relative ordering $w_{P_1} \gtrsim w_D > w_S$ of the
two-particle cluster DM weights, predicted based on the combination of
one-particle cluster DM weights, is confirmed numerically --- even
though the predicted weights are off.  This suggests that we might be
able to push a variant of the naive Operator-Based DM Truncation
Scheme through, by introducing additional rules on how one-particle
cluster DM eigenstates can be combined to give only the valid
many-particle cluster DM eigenstates.  We did not attempt to implement
such a truncation scheme, and test how badly its numerical accuracy is
affected by the ratio $w_{s_2}/w_{s_1}$ (by calculating the dispersion
relation, for example, as was done in Ref.~\onlinecite{cheong04b}),
because we feel that such a naive scheme was not in the spirit of
finding an appropriate `quasi-particle' description for the low-energy
excitations of our system of interacting spinless fermions given by
\eqref{eqn:strongintham}.

Finally, we realized that our numerical studies do not allow us to
conclude whether the cluster DM of the interacting system furnishes a
good `quasi-particle' description for the strongly-interacting system.
To be able to check this, we must be able to construct appropriate
superpositions of cluster DM eigenstates with different particle
numbers, so that the pseudo-Hamiltonian $\tilde{H}_C \sim \log\rho_C$ 
looks like the sum of a noninteracting pseudo-Hamiltonian
$\tilde{H}_{C,0}$ and a weak interacting pseudo-Hamiltonian
$\tilde{H}_{C,1}$.  Instead of a simple eigenvalue problem for the
cluster DM, the problem of discovering what `quasi-particle' operators
make up the cluster DM is an operator singular value decomposition
problem.  We will carefully define this operator singular value 
decomposition in a future paper, and describe how it can be applied to
the density matrix of two disjoint clusters to systematically extract
operators associated with independent quantum fluctuations within each
cluster, and their inter-cluster correlations.\cite{cheong07}

\begin{acknowledgments}
This research is supported by NSF grant DMR-0240953, and made use of
the computing facility of the Cornell Center for Materials Research (CCMR)
with support from the National Science Foundation Materials Research
Science and Engineering Centers (MRSEC) program (DMR-0079992).  SAC would
like to thank Garnet Chan for illuminating discussions on the numerical
implementation of the trace-down algorithm.
\end{acknowledgments}

\begin{appendix}

\section{Computational Complexity of Cluster Density-Matrix Calculation}
\label{appendix:computational}

In this appendix we compare the \emph{naive algorithm} and the
\emph{pre-sorted inner product algorithm}, based on 
\eqref{eqn:detailrhocmatrixelements} and
\eqref{eqn:presortinnerproduct} respectively, for numerically
computing the cluster density matrix, and determine their 
computational complexities.  To begin, we denote by $D(P)$ the size of 
the system Hilbert space with $P$ particles, $D_C(P_C)$ the size of 
the cluster Hilbert space with $P_C$ particles, and $D_E(P_C)$ the 
size of the environment Hilbert space with $P_E = P - P_C$ particles.
Noting that there can be no matrix elements between cluster 
configurations with different number of particles, we calculate each 
$P_C$ sector of the cluster DM separately.  To keep our notations 
compact, let us drop the $P$ and $P_C$ dependences in $D(P)$, and 
$D_{C,E}(P_C)$ respectively from this point onwards, and reinstate 
these dependences only when necessary.  Readers are referred to 
Appendix A in Ref.~\onlinecite{cheong05} for more technical details on 
the computational implementation of this trace-down calculation of the 
cluster DM.

In the naive algorithm based on \eqref{eqn:detailrhocmatrixelements},
the cluster DM matrix elements are computed by starting nested `for' 
loops in $\bl$ and $\bl'$, each running over $D_C$ indices.  For each 
pair of cluster configurations $\ket{\bl}$ and $\ket{\bl'}$, one would 
need to then determine which of the $P$-particle configurations 
$\ket{\bj}$ contain the two cluster configurations.  This involves 
running through the $D$ configurations in the system Hilbert space,
and for each configuration, comparing the $P$ occupied sites with the
$P_C$ occupied sites in the cluster configurations $\ket{\bl}$ and
$\ket{\bl'}$.  The computational effort incurred for this matching is thus
$O(D P)$.  Two vectors of indices, $\bi$, whose entries are the indices of
system configurations $\ket{\bj}$ giving cluster configuration $\ket{\bl}$,
and $\bi'$, whose entries are the indices of system configurations 
$\ket{\bj}$ giving cluster configuration $\ket{\bl'}$, are obtained.  The
lengths of these index vectors vary, but are of $O(D_E)$.  One can then
compare the two index vectors, at a computational cost of $O(D_E^2)$,
to find which pairs of system configurations giving cluster configurations
$\ket{\bl}$ and $\ket{\bl'}$ share the same environment configuration.
Following this, one can sum over the amplitude of such pairs, at a 
computational cost of $O(D_E)$, to obtain the cluster DM matrix element 
$\braket{\bl|\rho_C|\bl'}$.  For this naive algorithm, the net computational
effort is on the order of $D_C^2(D P + D_E^2 + D_E) \sim D_C^2 (D P + D_E^2)$.

In the pre-sorted inner-product algorithm based on
\eqref{eqn:presortinnerproduct}, we need to first run through $D$ 
system configurations to pre-sort the amplitudes in the ground-state 
wave function.  For each system configuration $\ket{\bj}$, we
determine at a computational cost of $P$ comparisons what cluster 
configuration $\ket{\bl}$ and environment configuration $\ket{\bm}$ it 
contains.  We then search through the cluster and environment Hilbert 
spaces to determine what the indices of $\ket{\bl}$ and $\ket{\bm}$ 
are in their respective Hilbert spaces, which incurs a computational 
effort on the order of $D_C P_C$ and $D_E P_E$ respectively.  Once 
these indices are determined, the amplitudes in the ground-state wave 
function are organized into a $D_C \times D_E$ matrix.  The net 
computational expenditure is thus on the order of $D(P + D_C P_C + D_E 
P_E) \sim D(D_C P_C + D_E P_E)$.  After sorting the ground-state wave 
function, we can then start nested
for loops in $\bl$ and $\bl'$, each running over $D_C$ indices, to
evaluate the matrix element $\braket{\bl|\rho_C|\bl'}$ as the inner product
between two vectors of length $D_E$.  This trace-down stage incurs a
computational cost of $O(D_C^2 D_E)$.  Overall, the computational cost is
on the order of $D(D_C P_C + D_E P_E) + D_C^2 D_E$.

For models allowing nearest-neighbor occupation, the system Hilbert space
is the direct product of the cluster Hilbert space and the environment
Hilbert space, i.e.~$D = D_C D_E$.  Since the number $P$ of particles is
small in any reasonable exact diagonalization, we can treat it as a 
$O(1)$ constant.  For small clusters, the size $D_C$ of the cluster Hilbert
space will also be small, so that the size $D_E$ of the environment Hilbert
space will be comparable in magnitude to $D$.  With these considerations,
we find that the computational cost for the naive algorithm is $O(D D_C^2 + 
D_C^2 D_E^2) \sim O(D^2)$, while the computational cost for the inner-%
product algorithm with pre-sorting is $O(D + D D_E + D_E) \sim O(D^2)$.
The efficiency of the two algorithms therefore depend on the
prefactors, the estimation of which requires more thorough analyses of
the two algorithms.

For a model such as \eqref{eqn:strongintham}, where nearest-neighbor
occupation is forbidden, the system Hilbert space is smaller than the
direct product of the cluster Hilbert space and the environment Hilbert
space, i.e.~$D < D_C D_E$.  Given again that $P$ and $D_C$ are small
numbers, the computational cost for the two algorithms are essentially
determined by the ratio $D_C D_E/D$.  This ratio is strongly dependent on
the dimensionality of the problem: the superfluous configurations
generated by the direct product of the cluster Hilbert space and the
environment Hilbert space are invalid because they contain nearest-neighbor
sites, right at the boundary between the cluster and its environment,
which are occupied.  In one dimension, the number of superfluous 
configurations is small, because the boundary between the cluster and
its environment consists only of two bonds, whatever the size $N_C$ of
the cluster.  In two dimensions, the boundary between the cluster and its
environment is a line cutting roughly $\sqrt{N_C}$ bonds.  The number of
superfluous configurations is then proportional to $\exp(\alpha_2\sqrt{N_C})$,
where $\alpha_2$ is a constant prefactor which depends on the shape of the
cluster.  In $d$ dimensions, the number of boundary bonds is on the order
of $N_C^{(d-1)/d}$, and the number of superfluous states is proportional
to $\exp(\alpha_d N_C^{(d-1)/d})$.  Therefore, in dimensions greater than
one, $D_C D_E$ become increasingly larger than $D$ as $N_C$ is increased,
and the inner-product algorithm with pre-sorting, which involves only one
power of $D_C D_E$, is more efficient than the naive algorithm, which 
involves $(D_C D_E)^2$.

\end{appendix}

\end{document}